\documentclass[aps,prl,twocolumn,preprintnumbers,superscriptaddress,amsmath,amssymb]{revtex4}

\usepackage{graphicx}
\usepackage{subfigure}
\usepackage{epsfig}
\usepackage{xcolor}
\usepackage{dcolumn}
\usepackage{bm}
\usepackage{ulem}	
\usepackage{color}
\usepackage{amsmath}
\usepackage{amsfonts}
\usepackage{amssymb}
\usepackage{lineno}
\usepackage{braket}

\begin{document}

\title{Distinguishing and controlling Mottness in 1$\bm T$-TaS$_2$ by ultrafast light}

\author{Changhua Bao}
\affiliation{State Key Laboratory of Low-Dimensional Quantum Physics and Department of Physics, Tsinghua University, Beijing 100084, P. R. China}

\author{Haoyuan Zhong}
\affiliation{State Key Laboratory of Low-Dimensional Quantum Physics and Department of Physics, Tsinghua University, Beijing 100084, P. R. China}

\author{Fei Wang}
\affiliation{State Key Laboratory of Low-Dimensional Quantum Physics and Department of Physics, Tsinghua University, Beijing 100084, P. R. China}

\author{Tianyun Lin}
\affiliation{State Key Laboratory of Low-Dimensional Quantum Physics and Department of Physics, Tsinghua University, Beijing 100084, P. R. China}

\author{Haoxiong Zhang}
\affiliation{State Key Laboratory of Low-Dimensional Quantum Physics and Department of Physics, Tsinghua University, Beijing 100084, P. R. China}

\author{Zhiyuan Sun}
\affiliation{State Key Laboratory of Low-Dimensional Quantum Physics and Department of Physics, Tsinghua University, Beijing 100084, P. R. China}

\author{Wenhui Duan}
\affiliation{State Key Laboratory of Low-Dimensional Quantum Physics and Department of Physics, Tsinghua University, Beijing 100084, P. R. China}
\affiliation{Frontier Science Center for Quantum Information, Beijing 100084, P. R. China}

\author{Shuyun Zhou}
\altaffiliation{Correspondence should be sent to syzhou@mail.tsinghua.edu.cn}
\affiliation{State Key Laboratory of Low-Dimensional Quantum Physics and Department of Physics, Tsinghua University, Beijing 100084, P. R. China}
\affiliation{Frontier Science Center for Quantum Information, Beijing 100084, P. R. China}

\date{\today}

\begin{abstract}

{\bf Distinguishing and controlling the extent of Mottness is important for materials where the energy scales of the onsite Coulomb repulsion U and the bandwidth W are comparable. Here we report the ultrafast electronic dynamics of  1$\bm T$-TaS$_2$ by ultrafast time- and angle-resolved photoemission spectroscopy.
A comparison of the electron dynamics for the newly-discovered intermediate phase (I-phase) as well as the low-temperature commensurate charge density wave (C-CDW) phase shows distinctive dynamics. While the I-phase is characterized by an instantaneous response and nearly time-resolution-limited fast relaxation ($\sim$200 fs), the C-CDW phase shows a delayed response and a slower relaxation (a few ps). Such distinctive dynamics reflect the different relaxation mechanisms and provide nonequilibrium signatures to distinguish the Mott insulating I-phase from the C-CDW band insulating phase. Moreover, a light-induced bandwidth reduction is observed in the C-CDW phase, pushing it toward the Mott insulating phase. Our work demonstrates the power of ultrafast light-matter interaction in both distinguishing and controlling the extent of Mottness on the ultrafast timescale.}

\end{abstract}

\maketitle

In condensed matter physics, the effect of electron-electron (el-el) correlation strongly depends on the relative energy scales of the onsite Coulomb repulsion U and the bandwidth W \cite{mott1949basis,imada1998metal}.  
When W $\gg$ U, the physics is dominated by electron hopping as well as electron-phonon (el-ph) interaction [Fig.~1(a)]. In contrast,  when W $\ll$ U, el-el correlation plays a critical role, which forbids double occupancy of electrons on a single site. This results in a Mott insulating state even at half filling [Fig.~1(b)], which could lead to exotic electronic states such as unconventional superconductivity upon doping \cite{damascelli2003angle,lee2006doping}. However, for insulators with comparable W and U [Fig.~1(c)], it is difficult to experimentally determine whether it is a Mott insulator from measurements in the equilibrium state. Finding an effective pathway to distinguish    and further control the extent of Mottness in the nonequilibrium state is therefore important.

Ultrafast pump-probe measurements provide opportunities for distinguishing the different types of insulators via their dynamics relaxation upon photoexcitation  \cite{hellmann2012time,wang2016using}. 
For band insulators, the nonequilibrium electronic dynamics typically involves a slow interband relaxation through radiation recombination, Auger recombination and carrier diffusion etc \cite{sundaram2002inducing}. 
In contrast, the electronic dynamics of Mott insulators is somewhat faster through el-el interaction \cite{okamoto2010ultrafast,strohmaier2010observation,lenarvcivc2013ultrafast,wegkamp2014instantaneous,mitrano2014pressure,tancogne2018ultrafast,li2022keldysh}. For example, photoexcitation of Mott insulator VO$_2$  leads to an instantaneous modification of the electronic correlation and collapse of the band gap \cite{wegkamp2014instantaneous}, and an ultrafast reduction of the Coulomb interaction is also expected for NiO \cite{tancogne2018ultrafast}.  In addition, a fast relaxation via annihilation of doublon (double carrier occupancy on a single site) and holon (no carrier occupancy) through  electron hopping \cite{strohmaier2010observation,lenarvcivc2013ultrafast}, as schematically illustrated in Fig.~1e, have also been reported for cuprate \cite{okamoto2010ultrafast}, ET-F$_2$TCNQ \cite{mitrano2014pressure} and Ca$_2$RuO$_4$ \cite{li2022keldysh}. Moreover, the comparable energy scales also make the material particularly sensitive to external perturbations such as ultrafast laser excitation \cite{stojchevska2014ultrafast}, and therefore, the light-matter interaction can also be potentially used as a control knob for tailoring the electronic structure by renormalizing U \cite{beaulieu2021ultrafast,baykusheva2022ultrafast} or W.

\begin{figure*}[htbp]
	\centering
	\includegraphics[width=16cm]{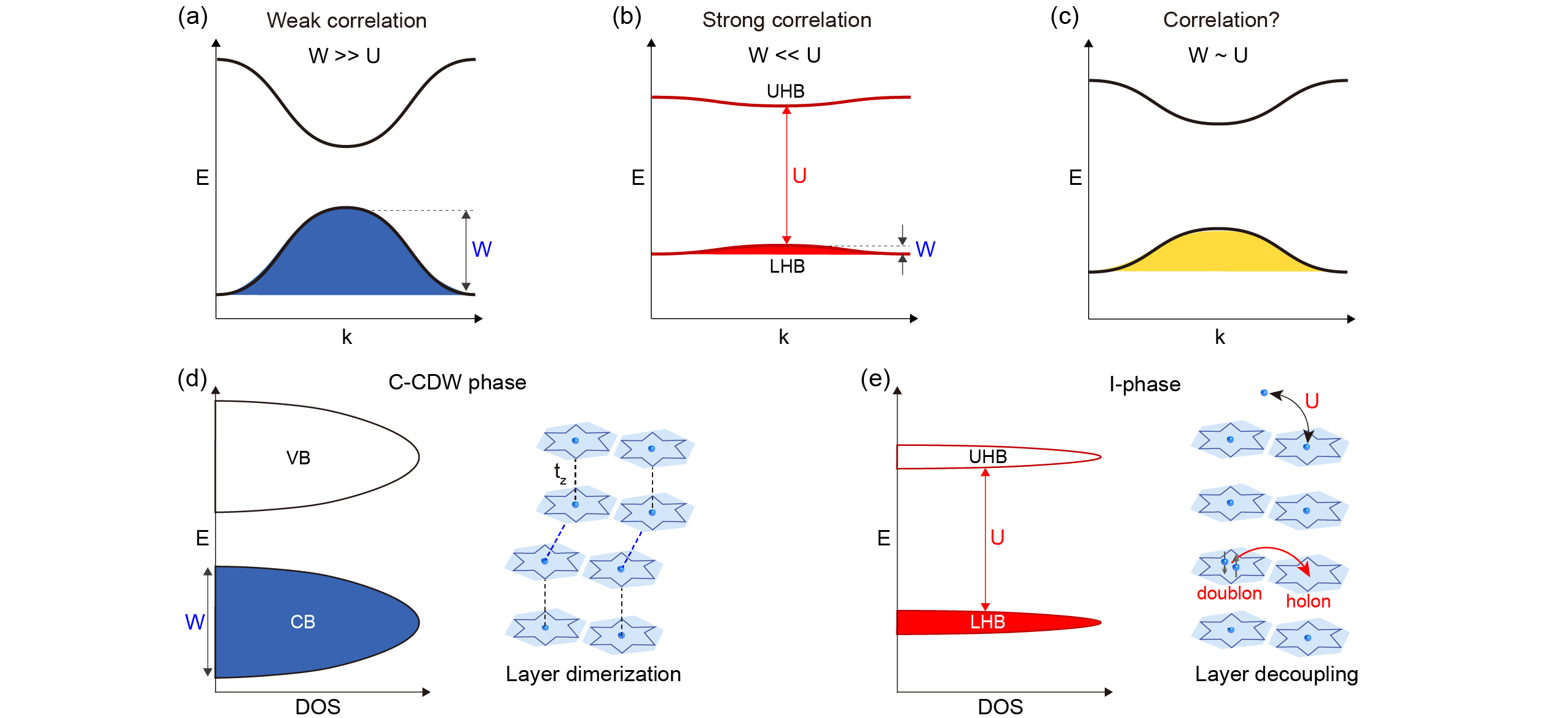}
	\caption{Classification of weak, strong and intermediate correlation according to the relative energy scales of W and U. (a)-(c) Schematic dispersions for weak, strong and intermediate correlation according to the relative energy scales of W and U. (d) Schematics for the energy diagram and layer dimerization in the C-CDW phase of 1$T$-TaS$_2$. (e) Schematics for the energy diagram and dominant interactions in the newly-discovered I-phase of 1$T$-TaS$_2$.}
\end{figure*}

Transition-metal dichalcogenide 1$T$-TaS$_2$ is a model material with similar energy scales of W and U. It shows rich charge density wave (CDW) phases at low temperatures \cite{wilson1975charge,scruby1975role}, which folds the bands into a smaller Brillouin zone and reduces W to be comparable to the effective U \cite{smith1985band,zwick1998spectral,kim1994observation,perfetti2005unexpected,ligges2018ultrafast}. Over the past few decades, the low-temperature commensurate CDW (C-CDW) phase had been considered as a Mott insulator because of the insulating behavior at half filling \cite{fazekas1979electrical,zwick1998spectral,sipos2008mott,law20171t,klanjvsek2017high}. However, recent experimental results suggest that the C-CDW phase shows a large bandwidth  \cite{ngankeu2017quasi,ritschel2018stacking,wang2020band}, which originates from the layer dimerization \cite{darancet2014three,ritschel2015orbital,butler2020mottness,stahl2020collapse,wang2020band}  and stacking order \cite{lee2019origin,butler2020mottness,lee2021distinguishing,wen2021roles,Werner2022PRL,nicholson2022modified}  [Fig.~1(d)]. The dimerization leads to unit cell doubling with two electrons occupying each unit cell (full band filling), suggesting that the C-CDW phase is likely a band insulator, although some extent of correlation or Mottness may also exist. More importantly, a new intermediate phase (I-phase) has been recently reported upon heating from the C-CDW phase into the triclinic CDW (T-CDW) phase \cite{wang2020band}. The I-phase shows a significantly reduced bandwidth, and the filling is reduced by half due to the removal of the layer dimerization [Fig.~1(e)], and thereby, it has been proposed to be a true Mott insulator.

Ultrafast time- and angle-resolved photoemission spectroscopy (TrARPES) is a powerful technique for revealing the different mechanisms underlying the ultrafast dynamics. While previous TrARPES works on  1$T$-TaS$_2$ have reported the carrier dynamics of the C-CDW phase, the room-temperature metallic phase and the pump-induced hidden phase \cite{perfetti2006time,perfetti2008femtosecond,petersen2011clocking,hellmann2012time,sohrt2014fast,avigo2016accessing,ligges2018ultrafast,avigo2019excitation,simoncig2021dissecting,maklar2022coherent,ren2022coexistence}, 
so far there is no report on the electronic dynamics of the newly-discovered I-phase yet. Here, we report the ultrafast electronic dynamics of the I-phase and provide nonequilibrium dynamical signatures to support that the I-phase is a Mott insulator while the C-CDW phase is a band insulator.  Moreover, a transient light-induced band flattening is observed in the C-CDW phase, pushing it toward the Mott insulating phase.

\begin{figure*}[htbp]
	\centering
	\includegraphics[width=15.5cm]{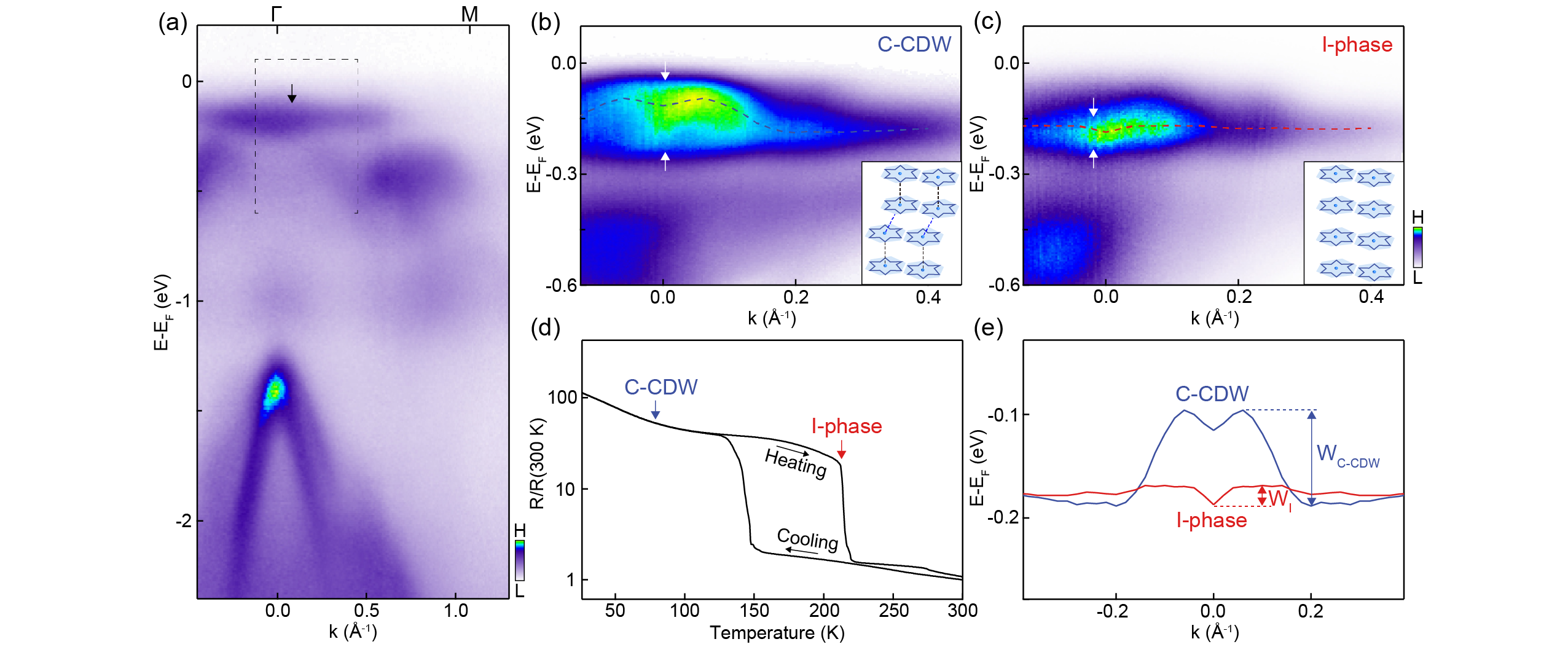}
	\caption{A comparison of the experimental electronic structures for C-CDW phase and I-phase. (a) Dispersion image of C-CDW phase by using Helium lamp source at 80 K. (b),(c) Zoom-in dispersion images measured near the $\Gamma$ point  (marked by dash rectangle in (a)) by using high-resolution laser source for C-CDW phase at 80 K (b) and I-phase at 210 K (c) upon heating. The insets are corresponding schematic crystal structures. (d) Normalized resistance as a function of temperature. (e) A comparison of dispersions extracted from the data shown in (b) and (c). The dispersions are symmetrized with respect to $\Gamma$ point.}
\end{figure*}

Figure 2 shows a comparison of the experimental electronic structures for the  C-CDW phase and the proposed Mott insulating I-phase. The large-range dispersion image of the C-CDW phase shows no band crossing at the Fermi energy $E_F$ in Fig.~2(a), in agreement with the insulating property from previous ARPES measurements \cite{zwick1998spectral,rossnagel2011origin}  as well as resistivity measurements in Fig.~2(d). 
Figures 2(b) and 2(c) show a comparison of the zoom-in dispersion images near the $\Gamma$ point with a high-resolution laser source for both C-CDW phase and I-phase. The comparison shows that the band near $E_F$ has a larger bandwidth and band broadening in the C-CDW phase [Fig.~2(b)], while the bandwidth is significantly reduced in the I-phase [Fig.~2(c)], which is in agreement with the proposed  Mott insulating phase by recent synchrotron-based ARPES study \cite{wang2020band}. The reduction of the bandwidth is more clearly observed from a comparison of the extracted dispersions shown in Fig.~2(e) (Fig.~S1 \cite{supp}). In particular, upon transition from the C-CDW phase to I-phase, the dispersion near the $\Gamma$ point shifts down while the dispersion at the momenta away from the $\Gamma$ point shifts up.  Such reduction of the bandwidth increases the extent of Mottness in the equilibrium state of the I-phase. In addition, it shows that the electronic structure near $\Gamma$ is more sensitive to the layer dimerization, while the electronic states away from $\Gamma$ are much less sensitive.

\begin{figure*}[htbp]
	\centering
	\includegraphics[width=16cm]{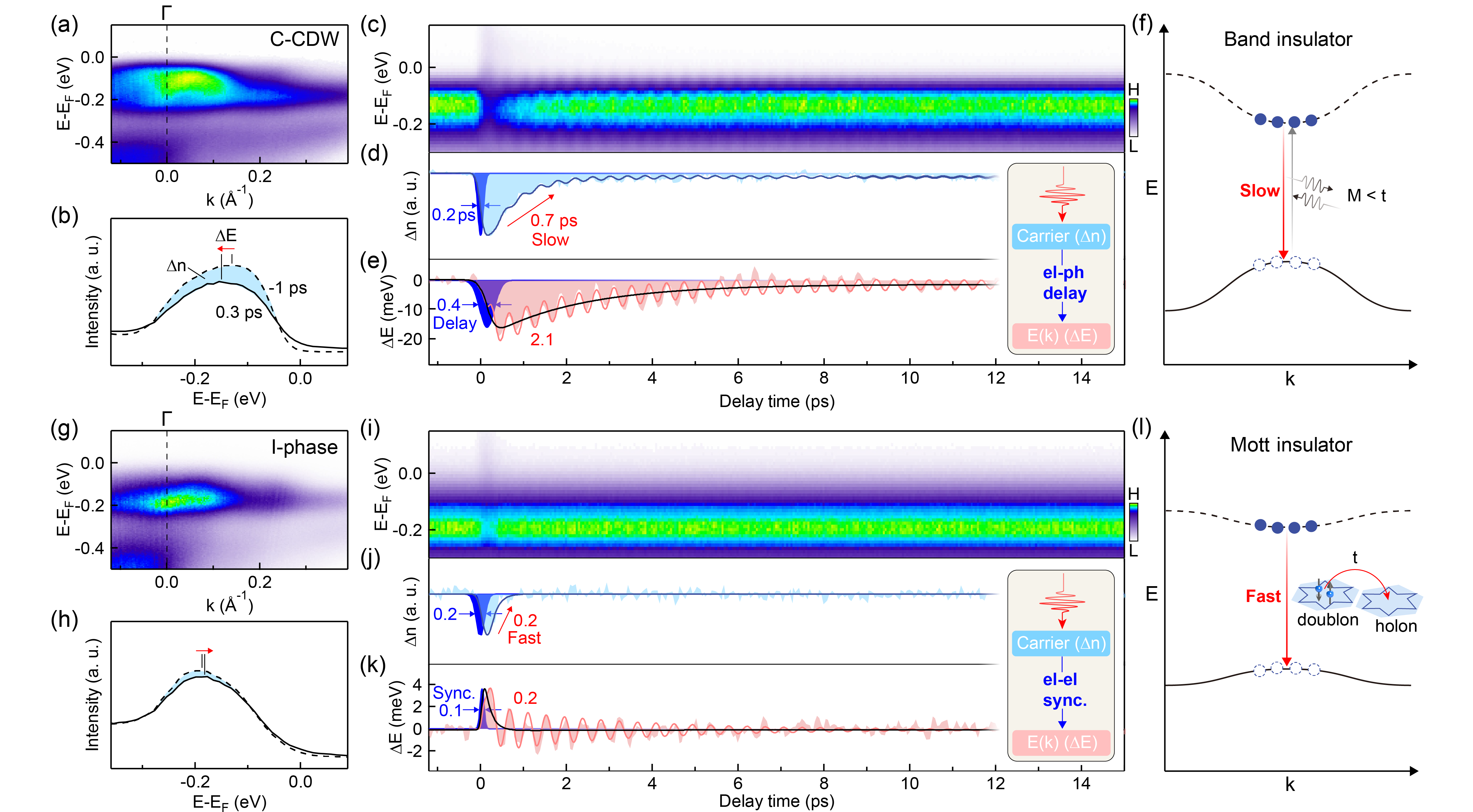}
	\caption{Distinguishing Mottness in the electronic dynamics. (a) Dispersion image of the C-CDW phase at 80 K before pumping. (b) EDCs at the $\Gamma$ point measured at $\Delta$t = -1 ps and 0.3 ps.  (c) Evolution of EDCs at the $\Gamma$ point as a function of delay time.  (d),(e) Extracted photocarrier density $\Delta$n and band shift $\Delta$E as a function of delay time. (f) A schematic of the slow relaxation in the C-CDW phase through radiation and reabsorption of photons.   (g)-(k) Similar data to (a)-(e) for the I-phase measured at 210 K. (l)  A schematic of the fast relaxation in the I-phase, which corresponds to the annihilation of a doublon and a holon.}
\end{figure*}

To reveal the nonequilibrium dynamics of these two phases in the time domain, TrARPES measurements have been performed with a weak pump of 0.3 mJ/cm$^2$ to avoid the pump-induced hidden phase \cite{ravnik2021time,gao2022snapshots}. We first focus on the electronic dynamics in the C-CDW phase [Fig.~3(a)]. A comparison of the energy distribution curves (EDCs) at the $\Gamma$ point before and after pumping shows that the peak position moves down in energy by 20 meV upon pumping in Fig.~3(b) indicating a light-induced band shift $\Delta$E as represented by the red arrow in Fig.~3(b). %We also note that upward or purely oscillating band shift has been reported in previous TrARPES studies \cite{perfetti2006time,perfetti2008femtosecond,petersen2011clocking,hellmann2012time,sohrt2014fast,simoncig2021dissecting}, possibly because of the different photon energy and temperature there and the uncertainty of momentum. 
In addition, the intensity also decreases upon pumping as represented by the blue area in Fig.~3(b), suggesting the emergence of photoexcited holes.  Therefore, the change in the intensity can be used to represent the amount of photoexcited holes ($\Delta$n). 

Figure 3(c) shows the continuous evolution of the EDCs with delay time, where clear oscillations in the peak position and peak intensity are observed. The extracted photoexcited carriers density $\Delta$n and light-induced band shift $\Delta$E are shown in Figs.~3(d) and 3(e), both of which show an oscillating behavior with a period of 410 fs (2.44 THz from the Fourier transformation shown in Fig.~S2 \cite{supp}) due to the CDW amplitude mode \cite{perfetti2006time}. Here we focus on the non-oscillating part including excitation (build-up) time as well as the relaxation time to search for signatures to distinguish the Mottness. First of all, $\Delta$n shows a build-up time of 0.2 ps, which is comparable to the experimental time resolution of 0.16 ps, while $\Delta$E shows a slower build-up time of 0.4 ps.  The delayed response between $\Delta$E and $\Delta$n is consistent with the timescale of el-ph interaction in 1$T$-TaS$_2$ \cite{eichberger2010snapshots}, suggesting that the evolution of the electronic band likely involves the el-ph interaction as illustrated in the inset of Fig.~3(d), and thus that the C-CDW phase is a band insulator. 
Secondly, the relaxation also gives some hints for the band insulator.  In particular, $\Delta$n shows a relaxation time of 0.7 ps with many periods of oscillation. One mechanism for this slow relaxation is through interband decay channels \cite{sundaram2002inducing} such as electron-hole recombination by emission of phonons or photons, as schematically illustrated in Fig.~3(f), which is typical for a band insulator. This process is slow because of the weak interband matrix element $M$ of the electron-boson interaction. 
The second possibility is that the relaxation comes from the recovery of the pump-suppressed interlayer CDW order (the layer dimerization), which is slow due to the involvement of the lattice degrees of freedom. We note that although the C-CDW phase seems to be a band insulator from the point of view in the time domain, some extent of correlation such as the Mottness may also exist, for example, a short-lived upper Hubbard band has been reported in some specific 1$T$-TaS$_2$ samples \cite{ligges2018ultrafast}.

In contrast to the C-CDW phase, the dynamics in the newly-discovered I-phase [Fig.~3(g)] is very different. The EDC analysis in Figs.~3(h) and 3(i) show a much weaker response upon pumping at the same pump fluence.  
Although the oscillation period of 430 fs (2.36 THz from the Fourier transformation in Fig.~S2 \cite{supp}) is overall similar to the C-CDW phase, both the build-up and relaxation time are faster than those in the C-CDW phase. Firstly, upon pumping, both $\Delta$n and $\Delta$E show a nearly resolution-limited response within 0.1--0.2 ps, and they are simultaneous without any delayed response [Figs.~3(j) and 3(k)], suggesting the band shift occurs on a much faster timescale and is driven by photoexcited carriers through el-el interaction, as illustrated in the inset of Fig.~3(j). 
Secondly, $\Delta$n relaxes within 0.2 ps, which is much faster than the 0.7 ps observed in the C-CDW phase. 
Such ultrafast relaxation time is in line with the reported fast electronic relaxation in other Mott insulators \cite{okamoto2010ultrafast,wegkamp2014instantaneous,mitrano2014pressure,lantz2017ultrafast,li2022keldysh,baykusheva2022ultrafast} and suggests that the relaxation is driven by annihilation of doublons and holons which is featured in Mott insulators. As illustrated in Fig.~3(l), the interband relaxation in a Mott insulator is intrinsically distinct from a band insulator due to the strong el-el correlation and corresponds to an electron hopping process from a double-occupied site (doublon) to the unoccupied site (holon) with a fast recombination rate exponentially dependent on the ratio of Hubbard U to exchange energy J \cite{strohmaier2010observation,lenarvcivc2013ultrafast}.
To summarize this part, the instantaneous band renormalization and photoexcited carriers with nearly time-resolution-limited excitation and relaxation times further support that the I-phase is a Mott insulating phase.

\begin{figure*}[htbp]
	\centering
	\includegraphics[width=16.5cm]{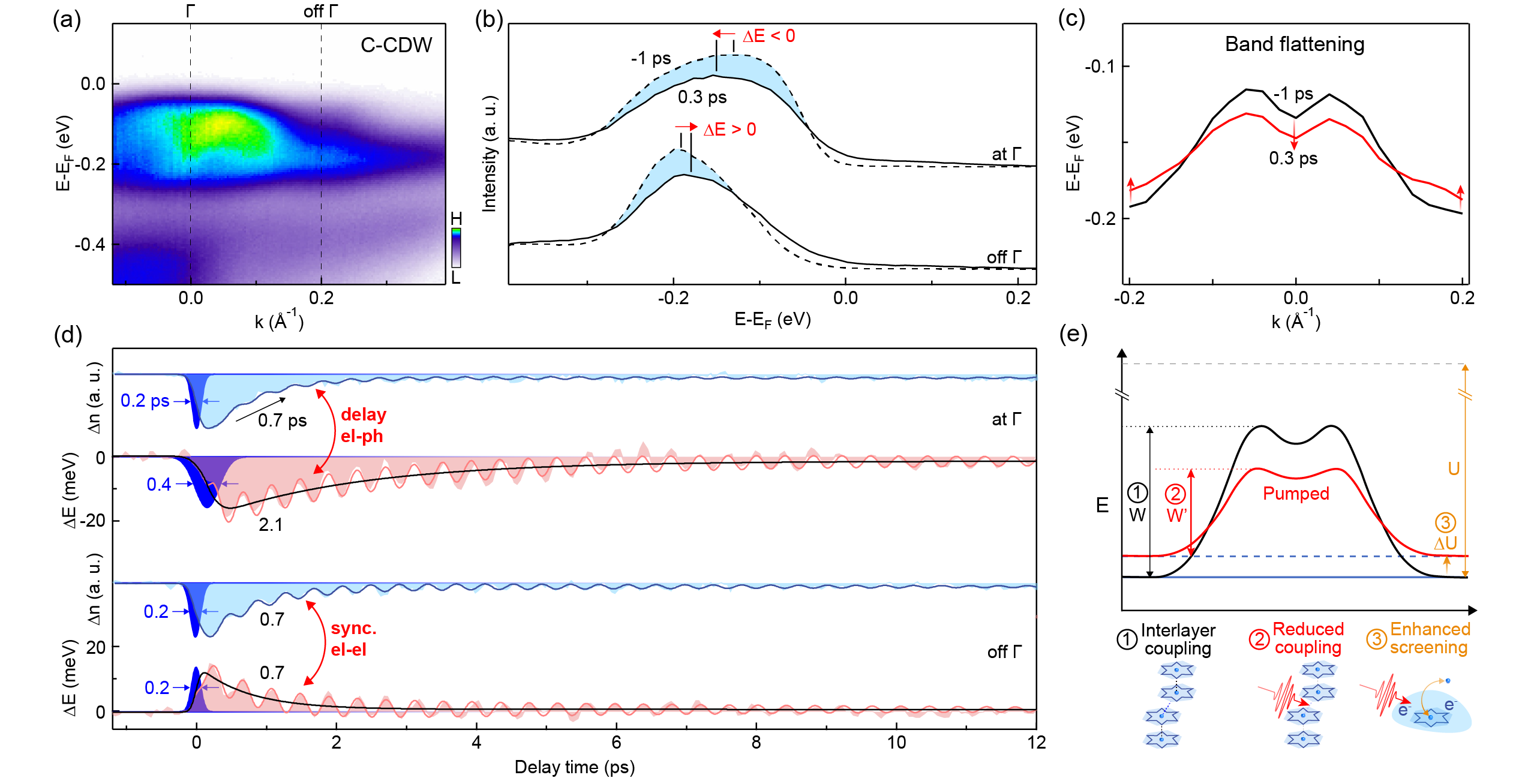}
	\caption{Light-induced band flattening in the C-CDW phase. (a) Dispersion image of the C-CDW phase. (b) EDCs at $\Gamma$ and off $\Gamma$ point (marked by dash lines in (a)) at $\Delta$t = -1 and 0.3 ps. (c) Extracted dispersions at $\Delta$t = -1 and 0.3 ps. (d) Extracted band shift $\Delta$E and photoexcited carrier density $\Delta$n as a function of delay time for $\Gamma$ and off $\Gamma$ point. (e) A schematic summary of the origin of momentum-dependent dynamics.}
\end{figure*}

Such ultrafast light-matter interaction can be used not only to distinguish the Mott insulating state, but also to control the extent of Mottness, which is supported by  the observed light-induced band flattening in the C-CDW phase. The light-induced band shift is found to be strongly momentum-dependent, in particular, the peak shifts down at the $\Gamma$ point, while it shifts up at a momentum away from the $\Gamma$ point, as shown in Fig.~4(b). 
The extracted dispersions before and after pumping further show that the opposite movements of the bands at and off the $\Gamma$ point lead to a reduction of the bandwidth from 80 to 50 meV, as shown in Fig.~4(c), demonstrating that ultrashort light pulses can be used to control the bandwidth, which is critical for determining the extent of Mottness. Here, the light-induced reduction of the bandwidth in the C-CDW phase pushes it toward the Mott insulating phase and increases the extent of Mottness.

To explore the underlying mechanism behind the exotic light-induced band flattening, Figure~4(d) shows the evolution of the $\Delta$E and $\Delta$n in the time domain both at and off the $\Gamma$ point, and a clear momentum dependence is observed in the dynamics. First, compared to the delayed evolution between $\Delta$E and $\Delta$n from el-ph coupling at the $\Gamma$ point, they are synchronized off the $\Gamma$ point, suggesting the band off the $\Gamma$ point is dominated by el-el interaction, in contrast to I-phase where synchronized $\Delta$E and $\Delta$n are observed both at  $\Gamma$ and off $\Gamma$ points (Fig.~S3 \cite{supp}). Second, $\Delta$E measured off the $\Gamma$ point shows an opposite sign from that measured at $\Gamma$ point. This can be understood by the photodoping induced screening as schematically illustrated in Fig.~4(e), which reduces U as reported in other correlated materials \cite{tancogne2018ultrafast,beaulieu2021ultrafast,baykusheva2022ultrafast} and leads to an upward band shift. The downward energy shift  at the $\Gamma$ point can be understood by the reduction of W due to light-induced weakening of the interlayer coupling, which is supported by the disappearance of the interlayer dimerization at a higher pump fluence revealed by ultrafast X-ray and electron diffraction measurements \cite{le2017stacking,stahl2020collapse} as well as rearrangement of the stacking order suggested by optical measurements \cite{li2020large}. Although U is also reduced by the photodoping effect, W shows a much larger reduction and longer lifetime, thus still pushing the system towards the Mott insulating phase \cite{supp}.

\begin{figure*}[htbp]
	\centering
	\includegraphics[width=15 cm]{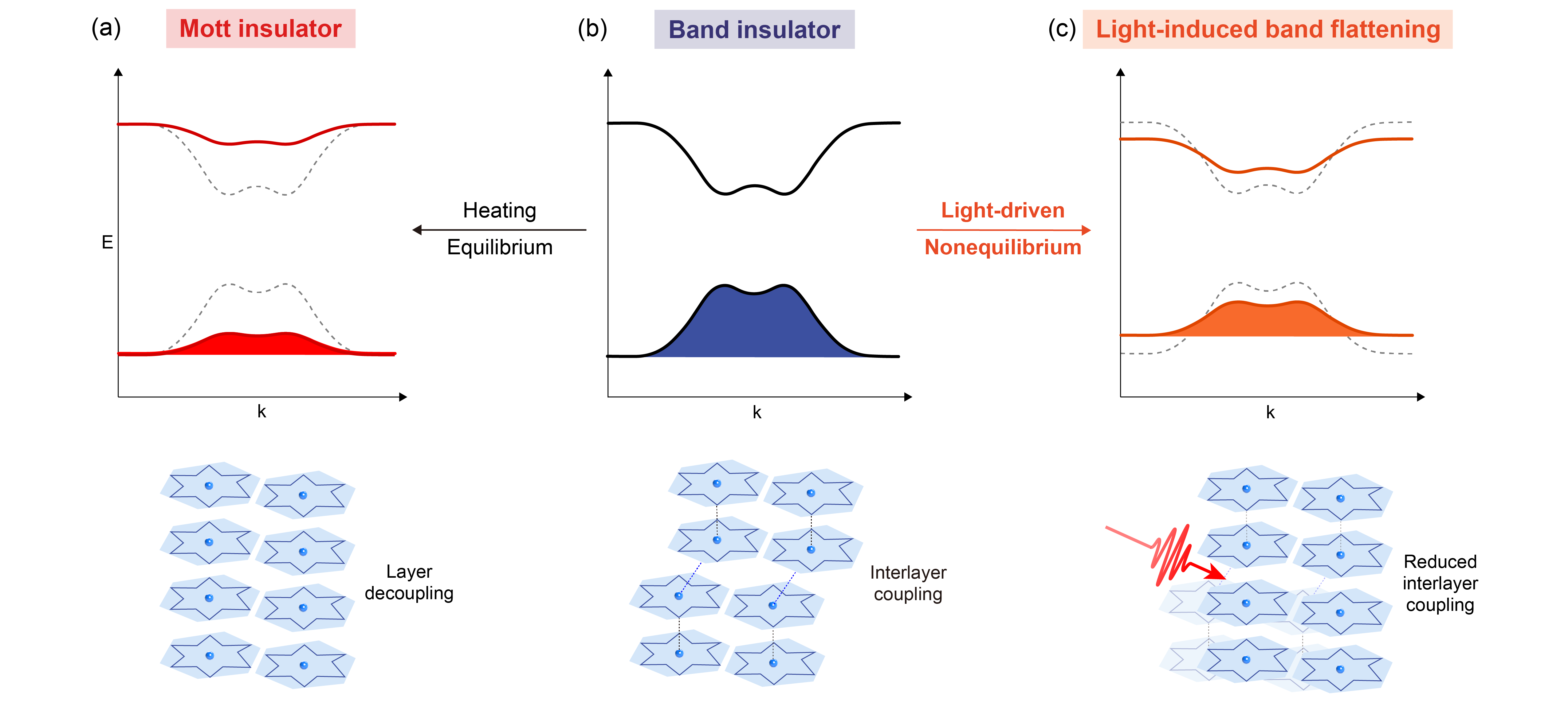}
	\caption{Band flattening upon photoexcitation and heating. (a) Mott insulating phase induced by heating upon interlayer decoupling. (b) The band insulating phase with dispersive band due to the interlayer coupling in the ground state. (c) The light-induced transient band flattening by reducing the interlayer coupling strength in the nonequilibrium state.}
\end{figure*}

In summary, our TrARPES measurements reveal the nearly time-resolution-limited excitation and ultrafast relaxation dynamics in the I-phase of 1$T$-TaS$_2$, in contrast to the delayed response and a slower relaxation in the C-CDW phase.  Such different responses suggest that the I-phase is dominated by el-el interaction, providing evidence for its Mott insulating nature from the nonequilibrium dynamics. In addition, thanks to the unique energy, momentum and time resolution of TrARPES measurements, a light-induced momentum-dependent band flattening is revealed in the C-CDW phase. The observed light-induced tuning of the electronic structure toward the Mott insulating phase is complimentary to heating-induced Mott insulating phase in the equilibrium state, while extending the control of the electronic structure to the ps timescale, as schematically summarized in Fig.~5. The light-induced reduction of the bandwidth is likely associated with the suppression of the interlayer dimerization, which was indicated by ultrafast diffraction and optical measurements \cite{le2017stacking,stahl2020collapse,li2020large}. Our work demonstrates that ultrafast light pulses can be used not only to distinguish, but also to control the Mottness on the ultrafast timescale, which is particularly useful for systems with similar energy scales of W and U.

\begin{acknowledgments}
\section*{ACKNOWLEDGMENTS}
This work was supported by the National Key R\&D Program of China  (No.~2021YFA1400100), the National Natural Science Foundation of China (No.~12234011, 92250305, 11725418, 11427903), and National Key R\&D Program of China (No.~2020YFA0308800). Changhua Bao is supported by Project funded by China
Postdoctoral Science Foundation (No.~2022M721886) and the Shuimu Tsinghua Scholar Program.

\end{acknowledgments}

\subsection{Conflict of interest}
The authors have no conflicts to disclose.

%\bibliography{reference}

\end{document}